# Learning to Localise Automated Vehicles in Challenging Environments using Inertial Navigation Systems (INS)

Uche Onyekpe[1, 2*], Vasile Palade[2] and Stratis Kanarachos[3]

[1] *Institute for Future transport and Cities, Coventry University, Gulson Road, Coventry, United Kingdom*
[2] *Research Center for Data Science, Coventry University, Gulson Road, Coventry, United Kingdom*
[3] *Faculty of Engineering, Coventry University, Gulson Road, Coventry, United Kingdom*
*onyekpeu@uni.coventry.ac.uk, ab8522@coventry.ac.uk, ab5839@coventry.ac.uk, ac0966@coventry.ac.uk*

***Abstract -*** *An algorithm based on Artificial Neural Networks is proposed in this paper to improve the accuracy of Inertial Navigation System (INS)/ Global Navigation Satellite System (GNSS) integrated navigation during the absence of GNSS signals. The INS which can be used to continuously position autonomous vehicles during GNSS signal losses around urban canyons, bridges, tunnels and trees, suffers from unbounded exponential error drifts cascaded over time during the integration of the gyroscope and double integration of the accelerometer to displacement. More so, the error drift is characterised by a pattern dependent on time. The Input Delay Neural Network (IDNN) has the ability to learn the error drift over time* [1] *and possesses the quality of being more computationally efficient than the Recurrent Neural Network (RNN), Long Short-Term Memory, and the Gated Recurrent Unit Network. Furthermore published literatures focus on travel routes which do not take complex driving scenarios into consideration, we therefore investigate in this paper the performance of the proposed algorithm on challenging scenarios, such as hard brake, roundabouts, sharp cornering, successive left and right turns and quick changes in vehicular acceleration across numerous test sequences. The results obtained show that the Neural Network-based approaches are able to provide up to 89.55 % improvement on the INS displacement estimation and 93.35 % on the INS orientation rate estimation.*

## *1. Introduction*

The safe navigation of autonomous vehicles and robots alike is dependent on fast and accurate positioning solutions. Autonomous vehicles are commonly localised within a lane using sensors such as cameras, LIDARS and radars, whilst road localisation is achieved through the use of information provided by a Global Navigation Satellite System (GNSS). There are however times when the LIDAR or/and camera might be unavailable for use [2], [3]. The GNSS which operates through the trilateration of signals obtained from at least three satellites is also unreliable. The accuracy of the GNSS deteriorates due to multi-path reflections and visibility issues under bridges and trees, in tunnels and urban canyons. An Inertial Navigation System (INS) can be used to estimate the position and orientation of the vehicle during the GNSS outage periods provided the availability of an initial orientation information [4]. Nevertheless, the INS, which is made up of sensors such as the accelerometer and gyroscope, suffers from an exponential error-drift during the double integration of the accelerometer's measurement to position and integration of the gyroscopes attitude rate to orientation [5]. These errors are cascaded unboundedly over time to provide a poor positioning solution within the navigation time window [5]. A common approach towards reducing the error drifts involves calibrating the INS periodically with the GNSS. The challenge therefore becomes one of accurately predicting the vehicles position in the absence of the GNSS signal needed for positioning and correction. Traditionally, Kalman filters are used in modelling the error between the GPS and INS positions [1], [6], [7]. However, they have limitations when modelling highly non-linear dependencies, stochastic relationships and non-Gaussian noise measurements [1].

Over the years, other methods based on artificial intelligence have been proposed by a number of researchers to learn the error-drift within the sensors [8]–[13]. The use of the sigma pi neural network on the positioning problem was explored by Malleswaran et al. in [14]. Noureldin et al. investigated the use of the Input Delay Neural Network (IDNN) to model the INS/GPS positional error [1]. A Multi-Layer Feed-Forward Neural Network (MFNN) was applied in [15] on a single point positioning INS/GPS integrated architecture. Rashad et al. in [16] employed the MFNN on an integrated tactical grade INS and Differential GPS architecture for a better position estimation solution. More so, Recurrent Neural Networks which are distinguishable from other neural networks due to their ability to make nodal connections in temporal sequences have been proven to model the time dependent error drift of the INS more accurately compared to other neural network techniques [17]. In [18], Fang et al. compared the performance of the LSTM algorithm to the Multi-Layer Neural Network and showed the superiority of the LSTM over the MLP.

---

[*] Corresponding author Uche Onyekpe: onyekpeu@uni.coventry.ac.uk

Similarly, in [2], Onyekpe et al. investigated the performance of the LSTM algorithm for high data rate positioning and showed in comparison to the IDNN, MLP and Kalman filter the superiority of the LSTM technique.

Nevertheless, we observe that despite the number of techniques investigated on the INS/GPS error drift modelling, there lacks an investigation into the performances of the techniques on complex driving scenarios and environments experienced in everyday driving. Such scenarios range from hard brakes on regular, wet or muddy roads to sharp cornering scenarios, heavy traffic, roundabouts etc. We thus set out to investigate the performances of artificial intelligence-based approaches on such complex driving environments and show that these scenarios prove rather more challenging for the INS. Furthermore, we propose an approach based on neural networks with inspiration drawn from the operation of the feedback control systems to improve the estimation of the neural network (NN) in tracking the vehicles displacement in these scenarios.

## *2. Problem Description and Formulation (INS Motion Model)*

Most of the research done on positioning do not take into consideration complex scenarios such as hard brake, sharp cornering or roundabouts. Hence, the evaluation of the performance of positioning techniques presented in most published works does not accurately reflect real-life vehicular driving experience. Moreover, as those complex scenarios present strong challenges for INS tracking, it seems essential for the reliability of the positioning methods that it be assessed under such scenarios.

- Hard brake – According to [19], hard brakes are characterised by a longitudinal deceleration of ≤−0.45 g. They occur when the brake pad of the vehicle has a large force applied to it. The sudden halt to the motion of the vehicle leads to a steep decline in the velocity of the vehicle, thus making it difficult to predict the vehicle coming to a stop and to track the vehicles motion thereafter. This scenario poses a major challenge to the displacement estimation of the vehicle.
- Sharp cornering and successive left right turns – The sudden and consecutive change in the direction of the vehicle also poses a challenge to the orientation estimation of the vehicle. The INS struggles to accurately capture the sudden sharp changes to the orientation of the vehicle as well as continuous consecutive changes to the vehicle orientation in relatively short periods of time.
- Changes in acceleration (Jerk) – The accuracy of the displacement estimation of the INS is affected by quick and varied changes to the acceleration of the vehicle within a short period of time. This is particularly a challenge as the INS struggles to capture the quick change in the vehicle's displacement thereafter.
- Roundabout – Roundabouts present a particular struggle due to its shape. The circular and unidirectional traffic flow makes it a challenge to track the vehicle's orientation and displacement particularly due to the continuous change in the vehicles direction whilst navigating the roundabout. Different roundabout sizes were considered in this study.

**2.1 INS/GPS motion model**

Tracking the position of a vehicle is usually done relative to a reference. The INSs measurements usually provided in the body (sensors) frame would need to be transferred to the navigation frame for tracking purposes [20]. In this study we adopt the North-East-Down convention in defining the navigation frame. The transformation matrix from the body frame to navigation frame is as shown in $Eq.\,1$

$$R^{nb} = \begin{bmatrix} cos\theta cos\Psi & -cos\phi\, sin\Psi + sin\phi\, sin\theta\, cos\Psi & sin\phi\, sin\Psi + cos\phi\, sin\theta\, cos\Psi \\ cos\theta\, sin\Psi & cos\phi\, cos\Psi + sin\phi\, sin\theta\, sin\Psi & -sin\phi\, cos\Psi + cos\phi\, sin\theta\, sin\Psi \\ -sin\theta & sin\phi cos\theta & cos\phi cos\theta \end{bmatrix} \quad (1)$$

Where $\phi$ is the roll, $\theta$ is the pitch and $\Psi$ is the yaw. However, as our study is limited to the one-dimensional tracking[1] of vehicles, $\phi$ and $\theta$ are thus considered to be zero thus the rotation matrix $R^{nb}$ becomes:

$$R^{nb} = \begin{bmatrix} cos\Psi & -sin\Psi & 0 \\ sin\Psi & cos\Psi & 0 \\ 0 & 0 & 1 \end{bmatrix} \quad (2)$$

The gyroscope measures the rate of change of attitude (angular velocity) in yaw, roll and pitch with respect to the inertial frame as expressed in the body frame [2], [20]. Giving initial orientation information, the attitude rate $\omega^b$ can be integrated to provide continuous information in the absence of the GNSS signal.

$$\Psi_{INS} = \Psi_0 + \int_{t-1}^{t} \omega_{INS}^b \quad (3)$$

---

[1] One directional tracking in the body frame corresponds to two-dimensional tracking in the navigation frame.

The accelerometer measures the specific force[2] $f^b$ on the sensor in the body frame and is as expressed in Eq. 4. Where $g^n$ represents the gravity vector, $R^{bn}$ is the rotation matrix from the navigation frame to the body frame, and $a^n$ denotes the linear acceleration of the sensor expressed in the navigation frame.

$$f^b = R^{bn}(a^n - g^n) \tag{4}$$

However, the accelerometer measurements at each time $t$ is usually corrupted by a bias $\delta^b_{INS}$ and noise $\varepsilon^b_a$ and is thus represented by $F^b_{INS}$ as shown in $Eq$ 5.

$$F^b_{INS} = f^b_{INS} + \delta^b_{INS} + \varepsilon^b_a \tag{5}$$

More so, the accelerometer's bias varies slowly with time and as such can be modelled as a constant parameter. Whilst the accelerometer's noise is somewhat characterised by a Gaussian distribution and modelled as $\varepsilon^b_a \sim N(0, \Sigma_a)$. Therefore, the specific measurement equation as expressed in Eq.4 can be expanded as shown below:

$$\text{from Eq.4, } a^b = f^b + g^b \tag{6}$$

$$F^b_{INS} = a^b_{INS} + \delta^b_{INS,a} + \varepsilon^b_a \tag{7}$$

$$a^b = F^b_{INS} - \delta^b_{INS,a} - \varepsilon^b_a \tag{8}$$

$$F^b_{INS} - \delta^b_{INS,a} = a^b + \varepsilon^b_a \tag{9}$$

$$\text{However, } a^b_{INS} = F^b_{INS} - \delta^b_{INS,a} \tag{10}$$

$$a^b_{INS} = a^b + \varepsilon^b_a \tag{11}$$

The vehicle's velocity in the body frame can be estimated through the integration of $Eq.$ 11 as shown below:

$$v^b_{INS} = \int_{t-1}^{t} (a^b) + \varepsilon^b_v \tag{12}$$

Through the double integration of $Eq.$ 11, the displacement of the vehicle in the body frame at time $t$ from $t-1$, $x^b_{INS}$, can also be determined as shown in $Eq.$ 13.

$$x^b_{INS} = \iint_{t-1}^{t} (a^b) + \varepsilon^b_x \tag{13}$$

Where $\delta^b_{INS,a}$ is the sensors bias in the body frame calculated as a constant parameter from the average reading of a stationary accelerometer ran for 20 minutes, $F^b_{INS}$ is the corrupted measurement of the accelerometer sensor at time t (sampling time), $g$ is the gravity vector and $\iint_{t-1}^{t} a^b, \int_{t-1}^{t} a^b$ and $a^b$ are the uncorrupted (true) displacement, velocity and acceleration respectively of the vehicle.

Thus, the Vehicle's true displacement is expressed as $x^b_{GPS} \approx \iint_{t-1}^{t} a^b$

Furthermore, $\varepsilon^b_x$ can be obtained by:

$$\varepsilon^b_x \approx x^b_{GPS} - x^b_{INS} \tag{14}$$

Using the North - East - Down (NED) system, the noise $\varepsilon^b_x$, displacement $x^b_{INS}$, velocity $v^b_{INS}$ and acceleration $a^b_{INS}$ of the vehicle in the body frame within the window $t-1$ to $t$ can be transformed to the navigation frame using $R^{nb}$ as shown in $Eqs.$ 15 to 19. However, the down axis is not considered in this study. More so, the window size in this study is defined a 1 second.

$$R^{nb}_{INS} \cdot a^b_{INS} \rightarrow a^n_{INS} \rightarrow a^b_{INS} \cdot \cos\Psi, a^b_{INS} \cdot \sin\Psi \tag{15}$$

$$R^{nb}_{INS} \cdot v^b_{INS} \rightarrow v^n_{INS} \rightarrow v^b_{INS} \cdot \cos\Psi_{INS}, v^b_{INS} \cdot \sin\Psi \tag{16}$$

$$R^{nb}_{INS} \cdot x^b_{INS} \rightarrow x^n_{INS} \rightarrow x^b_{INS} \cdot \cos\Psi, x^b_{INS} \cdot \sin\Psi \tag{17}$$

$$\text{Where: } R^{nb}_{INS} = \begin{bmatrix} \cos\Psi_{INS} & -\sin\Psi_{INS} & 0 \\ \sin\Psi_{INS} & \cos\Psi_{INS} & 0 \\ 0 & 0 & 1 \end{bmatrix} \tag{18}$$

---

[2] In the vehicle tracking application, the centrifugal acceleration is considered absorbed in the local gravity sector and the centrifugal acceleration considered negligible due to its small magnitude.

$$\begin{bmatrix} a_{INS}^{n,North} \\ a_{INS}^{n,North} \\ v_{INS}^{n,North} \\ v_{INS}^{n,North} \\ x_{INS}^{n,North} \\ x_{INS}^{n,East} \end{bmatrix} = \begin{bmatrix} a_{INS}^{b} \cdot \cos\Psi_{INS} \\ a_{INS}^{b} \cdot \sin\Psi_{INS} \\ v_{INS}^{b} \cdot \cos\Psi_{INS} \\ v_{INS}^{b} \cdot \sin\Psi_{INS} \\ x_{INS}^{b} \cdot \cos\Psi_{INS} \\ x_{INS}^{b} \cdot \sin\Psi_{INS} \end{bmatrix} \quad (19)$$

The vehicle's true displacement[3] $x_{GPS}^{b}$ is determined using the Vincenty's Inverse and applied according to [2] using the Python implementation [21].

### 2.2 Neural Network Localisation Model Set Up

We propose a displacement estimation model to minimise the effect of the noise in the accelerometer as illustrated on Figure 1. The proposed model which is analogous to the functioning of a closed loop or feedback control system operates in prediction mode by feeding back the output of the neural Network $x_{predicted}^{b}$ from window $t-1/t-2$ and the vehicles acceleration $a_{INS\,t}^{b}$ into the neural network to estimate the distance $x_{GPS\,t|t-1}^{b}$ covered by the vehicle as further illustrated on Figure 2.

However, as presented on Figure 1, during the training phase the NN is fed with the GNSS estimated displacement $x_{predicted\,t-1|t-2}^{b}$ rather than the output of the NN $x_{GPS\,t-1|t-2}^{b}$. Both models are structured this way due to the availability of the GPS signal during the training phase and its absence in the prediction phase. The output $x_{predicted\,t-1|t-2}^{b}$ is thus setup to mimic the functionality of the GNSS resolved displacement $x_{GPS}^{b}$ at window $t-1/t-2$ during the prediction operation.

Howbeit, as the prediction models input $x_{predicted\,t-1|t-2}^{b}$ never matches the training models input $x_{GPS\,t-1|t-2}^{b}$, the challenge becomes one of minimising the effect of the inexactness of $x_{predicted\,t-1|t-2}^{b}$ on the performance of the prediction model. We set about to address this by introducing a controlled random white Gaussian noise to the training models input $x_{GPS\,t-1|t-2}^{b}$, during the learning phase. This approach attempts to aid the NN to account for the impreciseness in the prediction output as an input. Figure 1 and 2 shows the training and prediction blocks of the displacement model, respectively.

Furthermore, we adopt a much simpler approach towards the estimation of the vehicles orientation rate as we found no performance benefit in utilising the feedback approach presented in the previous paragraphs. On the orientation rate estimation, the NN is made to learn the relationship between the yaw rate $\omega_{INS\,t}^{b}$ as provided by the gyroscope and the ground truth (yaw rate) $\omega_{GPS\,t}^{b}$ calculated from the information provided by the GPS.

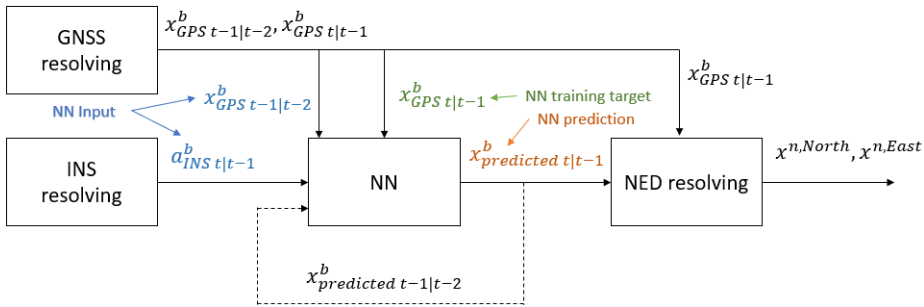

*Figure 1 Training Block of the proposed Displacement estimation model*

---

[3] $x_{GPS}^{b}$ is estimated as the distance between two points on the surface of the earth specified in longitude and latitude.

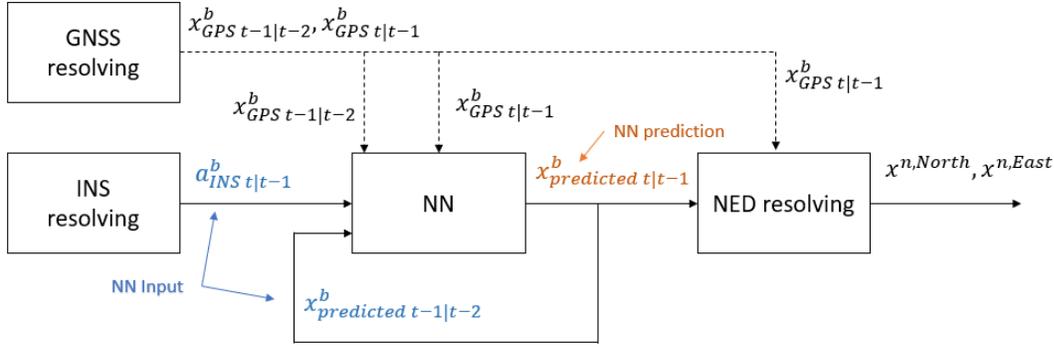

*Figure 2 Prediction Block of the proposed Displacement estimation model*

## 3. Experimental Setup & Data collection
### 3.1 Dataset
The IO-VNBD dataset consisting of 98 hours of driving data collected over 5700km of travel and characterised by diverse driving scenarios publicly available from https://github.com/onyekpeu/IO-VNBD is used [22]. The dataset captures information such as the vehicles longitudinal acceleration, yaw rate, heading, GPS co-ordinates (latitude longitude) at each time instance from the ECU of the vehicle with a sampling interval of 10 Hz. A Ford Fiesta Titanium is used for the data collection as shown on Figure 3. Tables 1, 7 and 8 presents the data subsets used in this study.

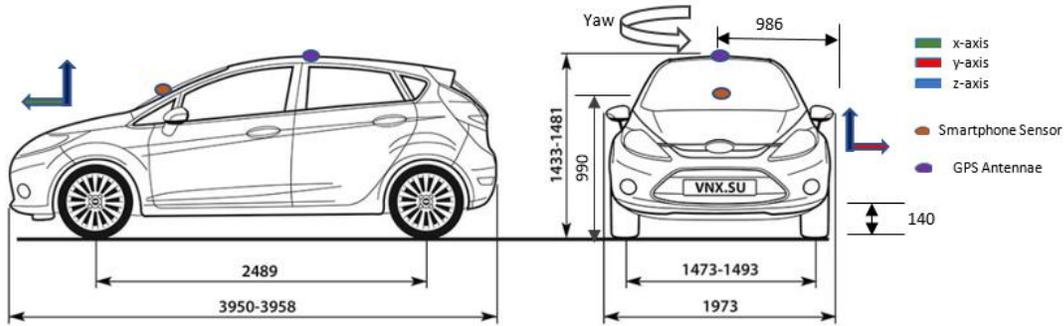

*Figure 3 Data collection vehicle showing sensor locations [22]*

*Table 1 IO-VNB data subsets used for the model training exercise [22]*

| IO-VNB Dataset | Features |
|---|---|
| **V-Vta1a** | Wet Road, Gravel Road, Country Road, Sloppy Roads, Round About (x3), Hard Brake on wet road, Tyre Pressure A |
| **V-Vta2** | Round About (x2), A Road (A511, A5121, A444), Country Road, Hard Brake, Tyre Pressure A |
| **V-Vta8** | Town Roads (Build-up), A-Roads (A511), Tyre Pressure A |
| **V-Vta10** | Round About (x1), A – Road (A50), Tyre Pressure A |
| **V-Vta16** | Round-About (x3), Hilly Road, Country Road, A-Road (A515), Tyre Pressure A |
| **V-Vta17** | Hilly Road, Hard-Brake, Stationary (No Mot ion), Tyre Pressure A |
| **V-Vta20** | Hilly Road, Approximate Straight-line t ravel, Tyre Pressure A |
| **V-Vta21** | Hilly Road, Tyre Pressure A |
| **V-Vta22** | Hilly Road, Hard Brake, Tyre Pressure A |
| **V-Vta27** | Gravel Road, Several Hilly Road, Potholes, Country Road, A-Road (A515), Tyre Pressure A |
| **V-Vta28** | Country Road, Hard Brake, Valley, A-Road (A515) |
| V-Vta29 | Hard Brake, Country Road, Hilly Road, Windy Road, Dirt Road, Wet Road, Reverse (x2), Bumps, Rain, B-Road (B5053), Country Road, U-Turn (x3), Windy Road, Valley, Tyre Pressure A |
| V-Vta30 | Rain, Wet Road, U-Turn (x2), A-Road (A53, A515), Inner Town Driving, B-Road (B5053), Tyre Pressure A |
| V-Vtb1 | Valley, rain, Wet-Road, Country Road, U-Turn (x2), Hard-Brake, Swift-Manoeuvre, A – Road (A6, A6020, A623, A515), B-Road (B6405), Round About (x3), day Time, Tyre Pressure A |
| V-Vtb2 | Country Road, Wet Road, Dirt Road, Tyre Pressure A |
| V-Vtb3 | Reverse, Wet Road, Dirt Road, Gravel Road, Night-time, Tyre Pressure A |
| V-Vtb5 | Dirt Road, Country Road, Gravel Road, Hard Brake, Wet Road, B Road (B6405, B6012, B5056), Inner Town Driving, A-Road, Motorway (M42, M1), Rush hour(Traffic) Round-About (x6), A-Road (A5, A42, A38, A615,A6), Tyre Pressure A |
| V-Vw4 | Round-About (x77), Swift-Manoeuvres, Hard-Brake, Inner City Driving, Reverse, A-Road, Motorway (M5, M40, M42), Country Road, Successive Left-Right Turns, Daytime, U-Turn (x3), Tyre Pressure D |
| V-Vw5 | Successive Left-Right Turns, Daytime, Sharp Turn Left/Right, Tyre Pressure D |
| V-Vw14b | Motorway (M42), Night-time, Tyre Pressure D |
| V-Vw14c | Motorway (M42), Round About (x2), A-Road (A446), Night-time, Hard Brake, Tyre Pressure D |
| V-Vfa01 | A-Road (A444), Round About (x1), B –Road (B4116) Day Time, Hard Brake, Tyre Pressure A |
| V-Vfa02 | B-Road (B4116), Round About (x5), A Road (A42, A641), Motorway (M1, M62) High Rise Buildings, Hard Brake, Tyre Pressure C |
| V-Vfb01a | City Centre Driving, Round-About (x1), Wet Road, Ring Road, Night, Tyre Pressure C |
| V-Vfb01b | Motorway (M606), Round-About (x1), City Roads Traffic, Wet Road, Changes in Acceleration in Short Periods of Time, Night, Tyre Pressure C |
| V-Vfb02b | Round About (x1), Bumps, Successive Left Right Turns, Hard-Brake (x7), Zig-zag (x6), Night, Tyre Pressure D |

## *3.2 Performance Evaluation Metric*

The performance of both the INS and NN based approaches are evaluated using the metrics defined below:

Cumulative Root Squared Error (CRSE) – The CRSE measures the cumulative root squared of the prediction error every second for the total duration of the GNSS outage defined as 10 seconds. It ignores the contributions of the negative sign of the error estimations enabling a better understanding of the performance of the positioning techniques.

$$CRSE = \sum_{t=1}^{N_t} \sqrt{e_{pred}^2}$$
(20)

Cumulative Absolute Error (CAE) – The CAE measures the absolute error of the prediction every second and summates the values throughout the duration of the GNSS outage, contrastingly to the CRSE signs are not ignored This tool is useful to better understand if the position technique is generally under or over predicting and how the prediction variance affects the overall positioning of the vehicle after the 10s outage period.

$$CAE = \sum_{t=1}^{N_t} e_{pred}$$
(21)

Average Error Per Second (AEPS) – The AEPS measures the average error of the prediction per second of the GNSS outage. It is useful in getting a microscopic view on the performance of the models.

$$AEPS = \frac{1}{N_t} \cdot e_{pred}$$
(22)

Mean ($\mu$) – The mean of the CRSE, CAE and AEPS across all sequences within each scenario is evaluated to reveal the average performance of the positioning technique in each scenario.

$$\mu_{CRSE} = \frac{1}{N_s}\sum_{i=1}^{N_s} CRSE, \quad \mu_{CAE} = \frac{1}{N_s}\sum_{i=1}^{N_s} CAE, \quad \mu_{AEPS} = \frac{1}{N_s}\sum_{i=1}^{N_s} AEPS$$
(23)

Standard Deviation ($\sigma$) – The standard deviation measures the variation of the CRSEs, CAEs and AEPSs of the sequences in each test scenario.

$$\sigma_{CRSE} = \sqrt{\frac{\sum(CRSE_i-\mu)^2}{N_s}}, \quad \sigma_{CAE} = \sqrt{\frac{\sum(CRSE_i-\mu)^2}{N_s}}, \quad \sigma_{AEPS} = \sqrt{\frac{\sum(CRSE_i-\mu)^2}{N_s}}$$
(24)

Where $N_t$ is GNSS outage length of 10 seconds, $t$ is the sampling period, $e_{pred}$ is the prediction error and $N_s$ is the total number of test sequences in each scenario.

Minimum (*min*) – The minimum metric informs of the minimum CRSE, CAE and AEPS across all sequences evaluated in each test scenario.

Maximum (*max*) – The maximum CRSE, CAE and AEPS provides information of the maximum CRSE across all sequences evaluated in each test scenario. It provides information on the possible accuracy of the positioning techniques in such scenarios. It is our rational that the *max* metric holds more significance compared to the $\mu$ and *min* as it captures the performance of the vehicle in each challenging scenario explored and further informs on the accuracy of the investigated techniques in each scenario.

## 3.3 Model Analysis for computational efficiency

The proliferation of Deep Learning and internet of things on low memory devices, increasing sensing and computing applications and capabilities promises to transform the performance of such devices on complex sensing tasks. The key impediment to the wider adoption and deployment of NN-based sensing application is their high computation cost. So therefore, there is the need to have a more compact parameterization of the neural Network models for easy deployment on embedded devices. To this end we evaluate the performance of the MLP, IDNN, RNN, GRU and LSTM on the round-about scenario across different parameterization for model efficiency.

From our study, we observe that the IDNN, RNN and GRU achieves a max CRSE orientation rate of 0.34 rad/s followed by the LSTM recording a max $CRSE$ of 0.35 rad/s whilst the MLP provided the worst performance of them all with a max $CRSE$ of 0.97 rad/s. In an almost similar fashion, the IDNN, RNN, LSTM and GRU obtain a max CRSE displacement of 17.96 m whilst the MLP obtains a max $CRSE$ of 315.77 m. However, as the IDNN is characterised by a significant lower number of parameters compared to the GRU, LSTM and RNN whilst providing similar CRSE scores across all NN studied and weight connections explored, we adopt it for use in learning the sensor noise in the accelerometer and gyroscope in this study. Table 2 shows the number of parameters characterizing each NN across the various weights investigated; 8, 16, 32, 64, 96, 128, 192, 256 and 320.

*Table 2 Number of parameters in each NN across various weighted connections*

| Number of weighted connections | Number of Trainable Parameters | | | | |
|---|---|---|---|---|---|
| | MLNN (2-Layer) | RNN (2-Layer) | GRU (2-Layer) | LSTM (2-Layer) | IDNN (2-Layer) |
| 8 | 33 | 65 | 185 | 245 | 65 |
| 16 | 97 | 225 | 657 | 873 | 161 |
| 32 | 321 | 833 | 2,465 | 3,281 | 449 |
| 64 | 1,153 | 3,201 | 9,537 | 2,705 | 1,409 |
| 128 | 4,353 | 12,545 | 37,505 | 49,985 | 4,865 |
| 192 | 9,601 | 28,033 | 83,905 | 111,841 | 10,369 |
| 256 | 16,897 | 49,665 | 148,737 | 198,273 | 17,921 |
| 320 | 26,241 | 77,441 | 232,001 | 309,281 | 27,521 |

Furthermore, we observe that the number of weighted parameters has little influence on the performance of the displacement and orientation rate estimation model. However, we notice that the number of time steps in the recurrent NN models (recurrent in both layer architecture and input structure such as the IDNN) significantly influences the accuracy of the model's prediction in both the orientation and displacement estimation. The performance of the IDNN across several time steps ranging from 2 to 14 are presented on Tables 3 and 4.

Table 3 Showing the performance evaluation based on the CRSE metric of the IDNN in each investigated scenario across several time steps on the orientation rate estimation

| Number of time steps | Motorway (rad/s) | Roundabout (rad/s) | Quick Changes in Vehicle Acceleration (rad/s) | Hard Brake (rad/s) | Sharp Cornering and Successive change in vehicle acceleration (rad/s) |
|---|---|---|---|---|---|
| 2 | 0.05 | 0.41 | 0.38 | 0.28 | 0.52 |
| 3 | 0.06 | 0.62 | 0.33 | 0.33 | 0.56 |
| 4 | 0.06 | 0.59 | 0.34 | 0.35 | 0.51 |
| 5 | 0.06 | 0.60 | 0.38 | 0.34 | 0.41 |
| 6 | 0.05 | 0.61 | 0.39 | 0.35 | 0.43 |
| 7 | 0.05 | 0.63 | 0.37 | 0.32 | 0.47 |
| 8 | 0.05 | 0.60 | 0.37 | 0.32 | 0.46 |
| 9 | 0.05 | 0.60 | 0.35 | 0.34 | 0.45 |
| 10 | 0.06 | 0.61 | 0.35 | 0.28 | 0.51 |
| 11 | 0.06 | 0.58 | 0.36 | 0.25 | 0.50 |
| 12 | 0.05 | 0.38 | 0.38 | 0.28 | 0.51 |
| 13 | 0.06 | 0.62 | 0.33 | 0.32 | 0.51 |
| 14 | 0.06 | 0.59 | 0.34 | 0.35 | 0.49 |

Table 4 Showing the performance evaluation based on the CRSE metric of the IDNN in each investigated scenario across several time steps on the displacement estimation

| NN | Number of time steps | Motorway (m) | Roundabout (m) | Quick Changes in Vehicle Acceleration (m) | Hard Brake (m) | Sharp Cornering and Successive change in vehicle acceleration (m) |
|---|---|---|---|---|---|---|
| IDNN | 2 | 651.41 | 702.17 | 571.99 | 648.57 | 425.56 |
|  | 4 | 616.60 | 655.22 | 546.76 | 580.01 | 373.03 |
|  | 6 | 610.61 | 599.22 | 524.41 | 577.19 | 346.92 |
|  | 8 | 592.27 | 595.09 | 474.55 | 557.50 | 292.78 |
|  | 10 | 3.60 | 17.96 | 8.71 | 15.80 | 14.55 |
|  | 12 | 3.23 | 19.52 | 8.62 | 19.36 | 14.43 |
|  | 14 | 3.63 | 20.53 | 9.58 | 20.45 | 12.71 |
| RNN | 2 | 17.11 | 55.58 | 33.30 | 64.09 | 42.64 |
|  | 4 | 7.87 | 47.92 | 22.45 | 51.81 | 25.95 |
|  | 6 | 7.28 | 29.10 | 16.56 | 25.28 | 22.58 |
|  | 8 | 3.83 | 21.08 | 10.27 | 16.39 | 14.02 |
|  | 10 | 3.60 | 17.96 | 8.71 | 15.80 | 14.55 |
|  | 12 | 3.23 | 19.52 | 8.62 | 19.36 | 14.43 |
|  | 14 | 3.63 | 20.53 | 9.58 | 20.45 | 12.71 |
| GRU | 2 | 21.19 | 51.54 | 36.66 | 62.04 | 45.82 |
|  | 4 | 25.01 | 42.62 | 30.11 | 45.05 | 26.19 |
|  | 6 | 20.24 | 33.72 | 24.64 | 23.87 | 22.20 |
|  | 8 | 11.33 | 24.14 | 15.16 | 17.59 | 15.47 |
|  | 10 | 3.60 | 17.96 | 8.71 | 15.80 | 14.55 |
|  | 12 | 3.23 | 19.52 | 8.62 | 19.36 | 14.43 |
|  | 14 | 3.63 | 20.53 | 9.58 | 20.45 | 12.71 |
| LSTM | 2 | 32.97 | 54.21 | 34.74 | 60.78 | 37.53 |
| | 4 | 18.20 | 41.62 | 26.73 | 51.94 | 28.21 |
| | 6 | 6.19 | 33.82 | 16.48 | 34.68 | 20.39 |
| | 8 | 4.12 | 21.75 | 11.49 | 16.18 | 13.8227 |
| | 10 | 3.60 | 17.96 | 8.71 | 15.80 | 14.55 |
| | 12 | 3.23 | 19.52 | 8.62 | 19.36 | 14.43 |
| | 14 | 3.63 | 20.53 | 9.58 | 20.45 | 12.71 |

### *3.4 Training of the IDNN model*

The displacement and orientation rate model were trained using the Keras-Tensorflow platform on the data subsets presented on Table 1 characterised by 800 minutes of drive time over a total travel distance of 760km. The models are trained using a mean absolute error loss function and an adamax optimiser with learning rates shown on Table 5. Furthermore, to avoid learning bias, all the features that were fed to the neural network were standardised between 0 and 1. Table 5 highlights the parameters characterising the training of the neural network approaches investigated.

*Table 5 Training parameters for the IDNN*

| Parameters | Displacement Estimation | Orientation rate Estimation |
|---|---|---|
| **Learning rate** | 0.004 | 0.001 |
| **Hidden units dropout rate** | 10% | 10% |
| **Time Steps (Sliding Window)** | See Table 4 | See Table 3 |
| **Hidden layers** | 2 | 2 |
| **Hidden neurons** | 32 per layer | 32 per layer |
| **Batch Size** | 256 | 256 |
| **Epochs** | 40 | 60 |

### *3.5 Testing*

The data subset used to investigate the performance of the INS and Neural Networks on the challenging scenarios are presented on Tables 6 and 7. Although then evaluated on complex scenarios as previously mentioned, the performance of the INS and NN modelling technique is first examined on the V-Vw12 dataset which presents a relatively easier scenario; an approximate straight line travel on the motorway. The evaluation on the latter scenario aims at gauging the performance of the technique in a relatively simpler driving situations. Nonetheless, the Motorway scenario could be challenging to track due to the large distance covered per second. In each scenario, the evaluation is conducted on several sequences of 10 seconds each with a prediction frequency of 1 second. GPS outages is assumed on the test scenarios, for the purpose of the investigation.

*Table 6 IO-VNB data test subset used in the less challenging scenario* [22]

| Scenario | IO-VNB Data subset | Total time driven, distance covered, velocity and acceleration |
|---|---|---|
| **Motorway** | V-Vw12 | 1.75 mins, 2.64 km, 82.6 - 97.4 km/hr, -0.06 - 0.07 g |

*Table 7 IO-VNB data test subset used in the challenging scenarios* [22]

| Challenging Scenarios | IO-VNB Data subset | Total time driven, distance covered, velocity and acceleration |
|---|---|---|
| **Roundabout** | V-Vta11 | 1.0 min, 0.92 km, 26.8 - 97.7 km/hr, -0.45 - 0.15 g |
| | V-Vfb02d | 1.5 mins, 0.84 km, 0.0 - 57.3 km/hr, -0.33 - 0.31 g |
| **Quick changes in acceleration** | V-Vfb02e | 1.6 mins, 1.52 km, 37.4 - 73.9 km/hr, -0.24 - 0.19 g |
| | V-Vta12 | 1.0 min, 1.27 km, 44.7 - 85.3 km/hr, -0.44 - 0.13 g |
| **Hard brake** | V-Vw16b | 2.0 mins, 1.99 km, 1.3 - 86.3 km/hr, -0.75 - 0.29 g |

| | V-Vw17 | 0.5 min, 0.54 km, 31.5 - 72.7 km/hr, -0.8 - 0.19 g |
| | V-Vta9 | 0.4 min, 0.43 km, 48.9 - 87.7 km/hr, -0.6 - 0.14 g |
| **Sharp cornering and successive left and right turns** | V-Vw6 | 2.1 mins, 1.08 km, 3.3 - 40.7 km/hr, -0.34 - 0.26 g |
| | V-Vw7 | 2.8 mins, 1.23 km, 0.4 - 42.2 km/hr, -0.37 - 0.37 g |
| | V-Vw8 | 2.7 mins, 1.12 km, 0.0 - 46.4 km/hr, -0.37 - 0.27 g |

## 4. Results and Discussion

The performance of the dead reckoned INS (INS DR) and proposed NN approaches are analysed comparatively across several GPS outage simulated sequences each of 10 s length. The positioning techniques are first analysed on a less challenging scenario involving vehicle travel on an approximate straight line on the motorway. Further analysis is then done on more challenging scenarios such as, hard brake, roundabouts, quick changes in acceleration and sharp cornering and successive left and right turns using the performance metrics defined in *section* 3.2.

### 4.1 Motorway Scenario

In evaluating the performance of the INS and the NN approaches on vehicular motion tracking, both techniques are investigated on a less challenging trajectory characterised by an approximate straight-line drive on the motorway. The results so gotten as shown on Table 8 shows that across all 9 test sequences, the INS records its best and average displacement and orientation rate CRSE of 1.63m, 15.01m, and 0.08 rad/s, 0.13 rad/s respectively. However, we observe that the NN outperforms the INS significantly on the displacement and orientation rate estimation across all metrics as illustrated on Figure 4. Comparatively, the average displacement and orientation rate CRSE of the NN approach is recorded as 0.84m, and 0.02 rad/s. Essentially, this shows that the NN maintained an average CRSE estimation error of 1.93m and 0.04 rad/s and CAE average estimation error of 0.96m and 001 rad/s after about 251m of travel. Being the lowest errors across all scenarios evaluated as shown on Figures 4, 14, 15 and 16 we can infer that this was the least challenging scenario due to minimal accelerations and directional change. Furthermore, the reliability of the NN in consistently tracking the vehicles motion with such accuracy is highlighted by its low standard deviation of 0.84. Figure 6b shows the trajectory of the vehicle along the motorway.

*Table 8 showing the performance of the IDNN and INS DR on the motorway scenario*

| **Displacement** | | **IDNN (m)** | **INS DR (m)** | **IDNN (rad/s)** | **INS DR (rad/s)** |
|---|---|---|---|---|---|
| **CRSE** | *max* | 3.23 | 30.11 | 0.05 | 0.21 |
| | *min* | 0.84 | 1.63 | 0.02 | 0.08 |
| | $\mu$ | 1.93 | 15.01 | 0.04 | 0.13 |
| | $\sigma$ | 0.84 | 9.12 | 0.01 | 0.03 |
| **CAE** | *max* | 2.56 | 30.11 | 0.02 | 0.13 |
| | *min* | 0.19 | 0.04 | 0.00 | 0.04 |
| | $\mu$ | 0.96 | 13.33 | 0.01 | 0.10 |
| | $\sigma$ | 0.84 | 10.17 | 0.01 | 0.03 |
| **AEPS (/s)** | *max* | 0.06 | 0.30 | 0.00 | 0.00 |
| | *min* | 0.00 | 0.01 | 0.00 | 0.00 |
| | $\mu$ | 0.02 | 0.15 | 0.00 | 0.00 |
| | $\sigma$ | 0.02 | 0.10 | 0.00 | 0.00 |
| **Total Distance Covered by vehicle** | *max* | 268.40 | | | |
| | *min* | 234.47 | | | |
| | $\mu$ | 251.29 | | | |
| **Number of Sequences evaluated** | | 9 | | | |

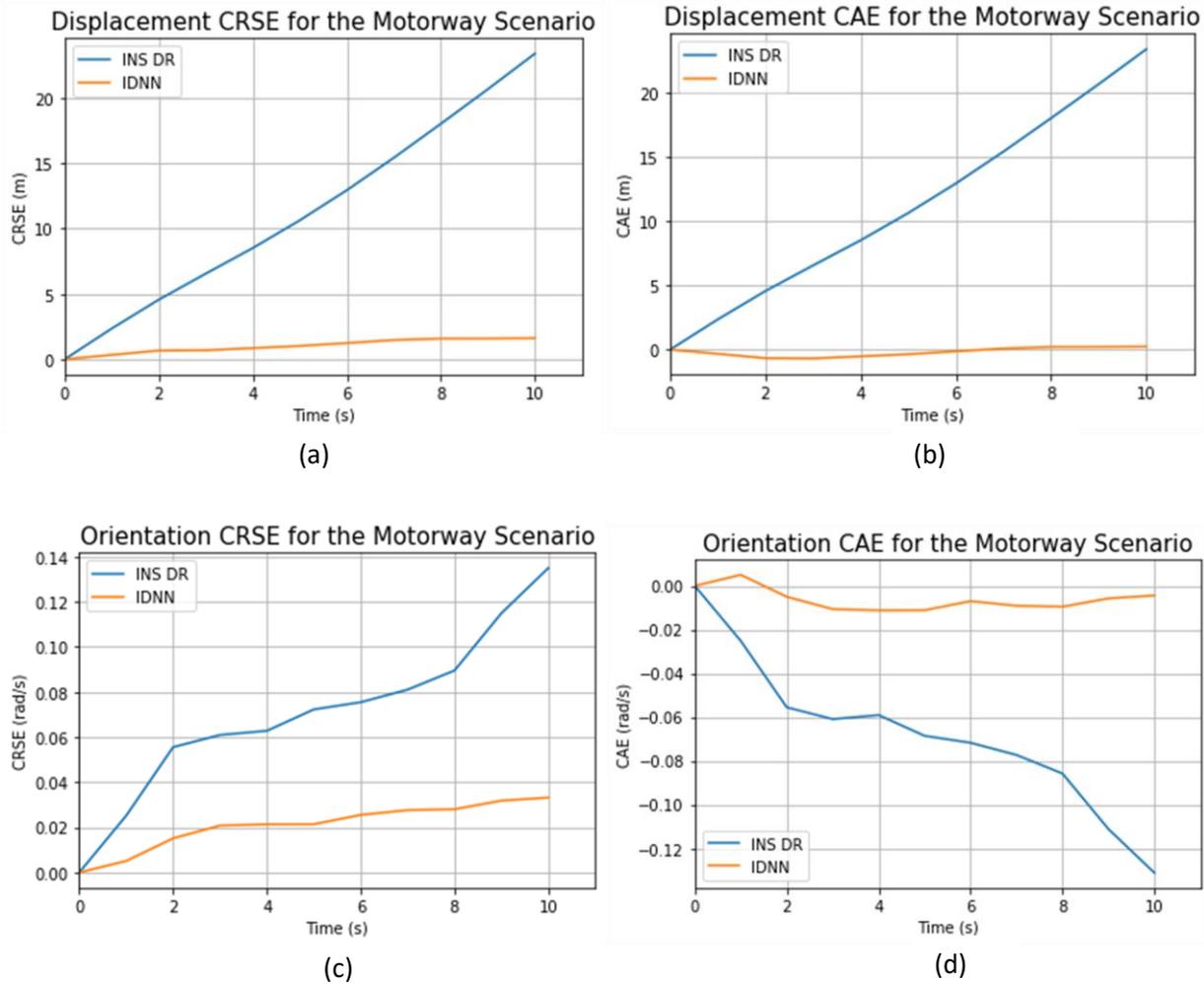

*Figure 4 showing the evolution of the estimation error over time in the motorway scenario based on the (a) displacement CRSE, (b) displacement CAE, (c) orientation rate CRSE and (d) orientation rate CAE*

### *4.2 Roundabout Scenario*

The roundabout scenario is one of the most challenging for both the vehicular displacement and orientation rate estimation. The difficulty encountered by the INS and NN in accurately tracking the vehicle's motion is graphically shown in Figures 5 with a comparative illustration to other investigated scenarios presented on Figures 14-16. The results so obtained as presented on Table 9 shows that the NN recorded a lower displacement (maximum CRSE and CAE of 17.96 m and 16.80 m) than the INS (maximum CRSE and CAE displacement of 171.92m) as well as a lower orientation rate (maximum CRSE and CAE of 0.38 rad/s and 0.04 rad/s respectively) compared to the INS (maximum CRSE and CAE of 5.71 rad/s and 2.14 rad/s respectively). The relatively higher standard deviation across all analysed metrics is evidence that the NN is able to less consistently track the vehicles position rate on the roundabout scenario but more consistently on other investigated scenarios. The roundabout scenario study was carried out across 11 test sequences over a maximum travel distance of approximately 197 m. Figure 6b shows a sample trajectory of the vehicle on the roundabout scenario analysis. The distribution of the gyroscopes measurement over time is presented on Figure 9a.

Table 9 showing the performance of the IDNN and INS DR on the roundabout scenario

| Displacement | | IDNN (m) | INS DR (m) | IDNN (rad/s) | INS DR (rad/s) |
|---|---|---|---|---|---|
| CRSE | max | 17.96 | 171.92 | 0.38 | 5.71 |
| | min | 2.36 | 19.42 | 0.02 | 0.17 |
| | $\mu$ | 8.63 | 78.32 | 0.17 | 1.48 |
| | $\sigma$ | 5.51 | 52.33 | 0.13 | 1.77 |
| CAE | max | 16.80 | 171.92 | 0.05 | 2.14 |
| | min | 0.5 | 19.42 | 0.00 | 0.03 |
| | $\mu$ | 5.47 | -57.60 | 0.01 | 0.62 |
| | $\sigma$ | 5.35 | 54.20 | 0.02 | 0.76 |
| AEPS (/s) | max | 0.34 | 1.76 | 0.01 | 0.08 |
| | min | 0.00 | 0.23 | 0.00 | 0.00 |
| | $\mu$ | 0.10 | 0.89 | 0.00 | 0.02 |
| | $\sigma$ | 0.10 | 0.49 | 0.00 | 0.03 |
| Total Distance Covered by vehicle | max | 196.71 | | | |
| | min | 19.89 | | | |
| | $\mu$ | 104.93 | | | |
| Number of Sequences evaluated | | 11 | | | |

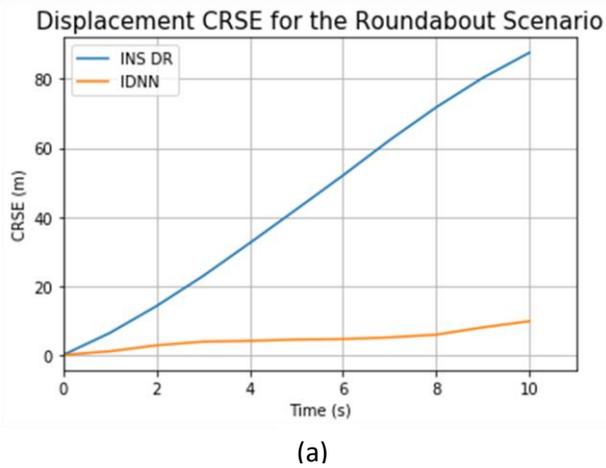

(a)

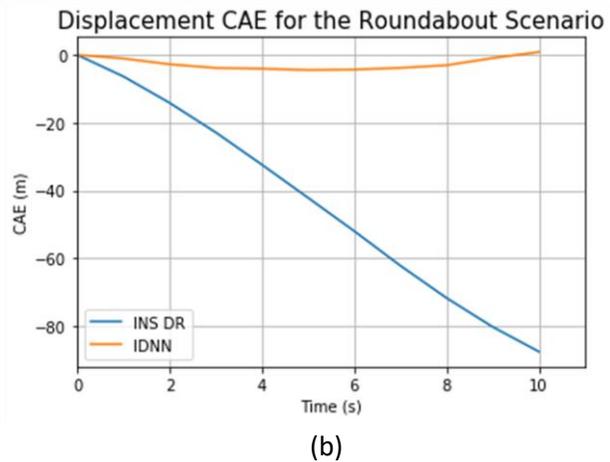

(b)

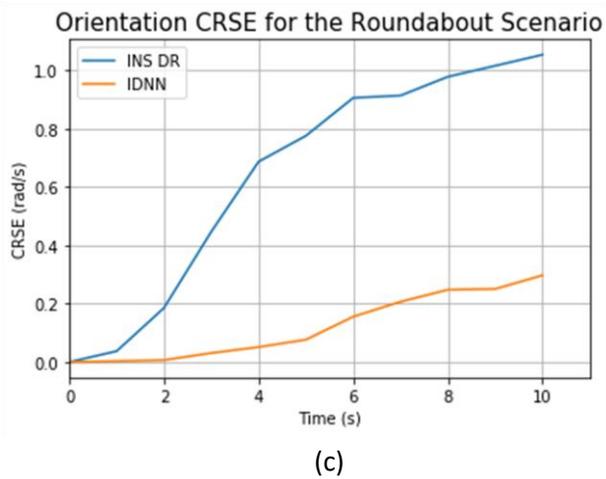

(c)

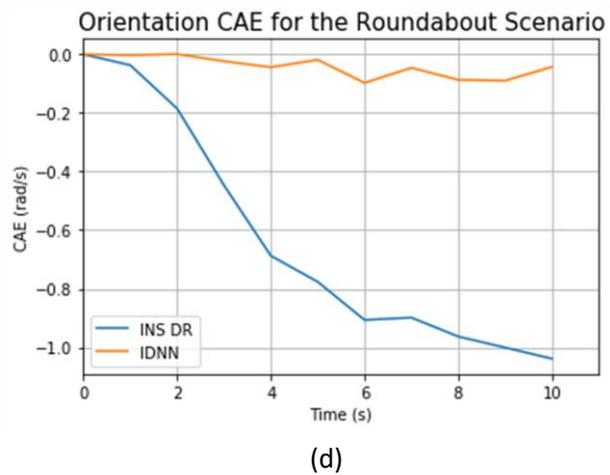

(d)

Figure 5 showing the evolution of the estimation error over time in the roundabout scenario based on the (a) displacement CRSE, (b) displacement CAE, (c) orientation rate CRSE and (d) orientation rate CAE

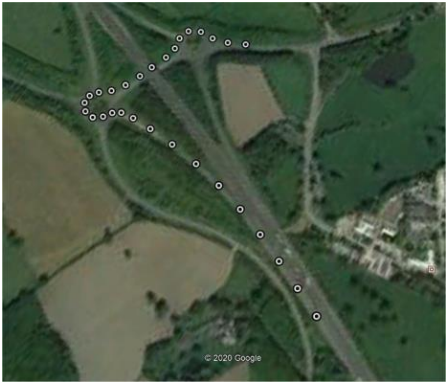 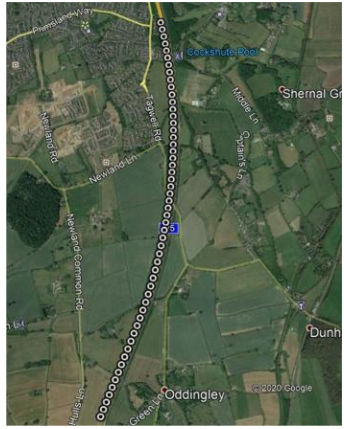

(a)                          (b)

*Figure 6 Sample trajectory of the (a) V-Vta11 roundabout data subset of the IO-VNBD and (b) V-Vw12 motorway data subset of the IO-VNBD*

## 4.3 Quick Changes in Vehicles Acceleration Scenario

The results presented on Table 10 illustrates the performance of the NN based approach over the INS in the quick changes in acceleration scenario. From observation it can be seen that the NN significantly outperforms the INS across all metrics employed with a maximum CRSE of 8.62 meters, 0.33 rad/s for the INS against 79.05 m and 0.67 rad/s for the INS over a maximum distance of approximately 220 m covered. This shows, as expected, that the INS and NN found it more challenging to estimate the displacement of the vehicle compared to the orientation rate. On other metrics the NN obtains an average CRSE, CAE and AEPS of 5.30 m, 0.93 m and 0.05m/s compared to that of the INS recorded as 38.92 m, 26.23 m and 0.43 m/s$^2$ across all 13 sequences evaluated. Figures 7 graphically illustrates the evolution of the error in a sample sequence across the CRSE and CAE metrics. A comparison of the performance of both approaches across all scenarios investigated are further presented in the Figures 14-16. Figure 9b shows a sample of the distribution of the vehicles acceleration over time in the Quick changes in acceleration scenario.

*Table 10 showing the performance of the IDNN and INS DR on the quick changes in acceleration scenario*

| Displacement | | IDNN (m) | INS DR (m) | IDNN (rad/s) | INS DR (rad/s) |
|---|---|---|---|---|---|
| **CRSE** | max | 8.62 | 79.05 | 0.33 | 0.67 |
| | min | 2.37 | 17.72 | 0.03 | 0.15 |
| | μ | 5.30 | 38.92 | 0.16 | 0.28 |
| | σ | 2.19 | 16.72 | 0.10 | 0.13 |
| **CAE** | max | 3.95 | 79.05 | 0.05 | 0.65 |
| | min | 2.37 | 17.72 | 0.00 | 0.00 |
| | μ | 0.93 | 26.23 | 0.02 | 0.18 |
| | σ | 2.83 | 32.44 | 0.01 | 0.17 |
| **AEPS (/s)** | max | 0.15 | 0.91 | 0.01 | 0.02 |
| | min | 0.01 | 0.02 | 0.00 | 0.00 |
| | μ | 0.05 | 0.43 | 0.00 | 0.00 |
| | σ | 0.03 | 0.25 | 0.00 | 0.00 |
| **Total Distance Covered** | max | 220.08 | | | |
| | min | 137.72 | | | |
| | μ | 168.62 | | | |
| **Number of Sequences evaluated** | | 13 | | | |

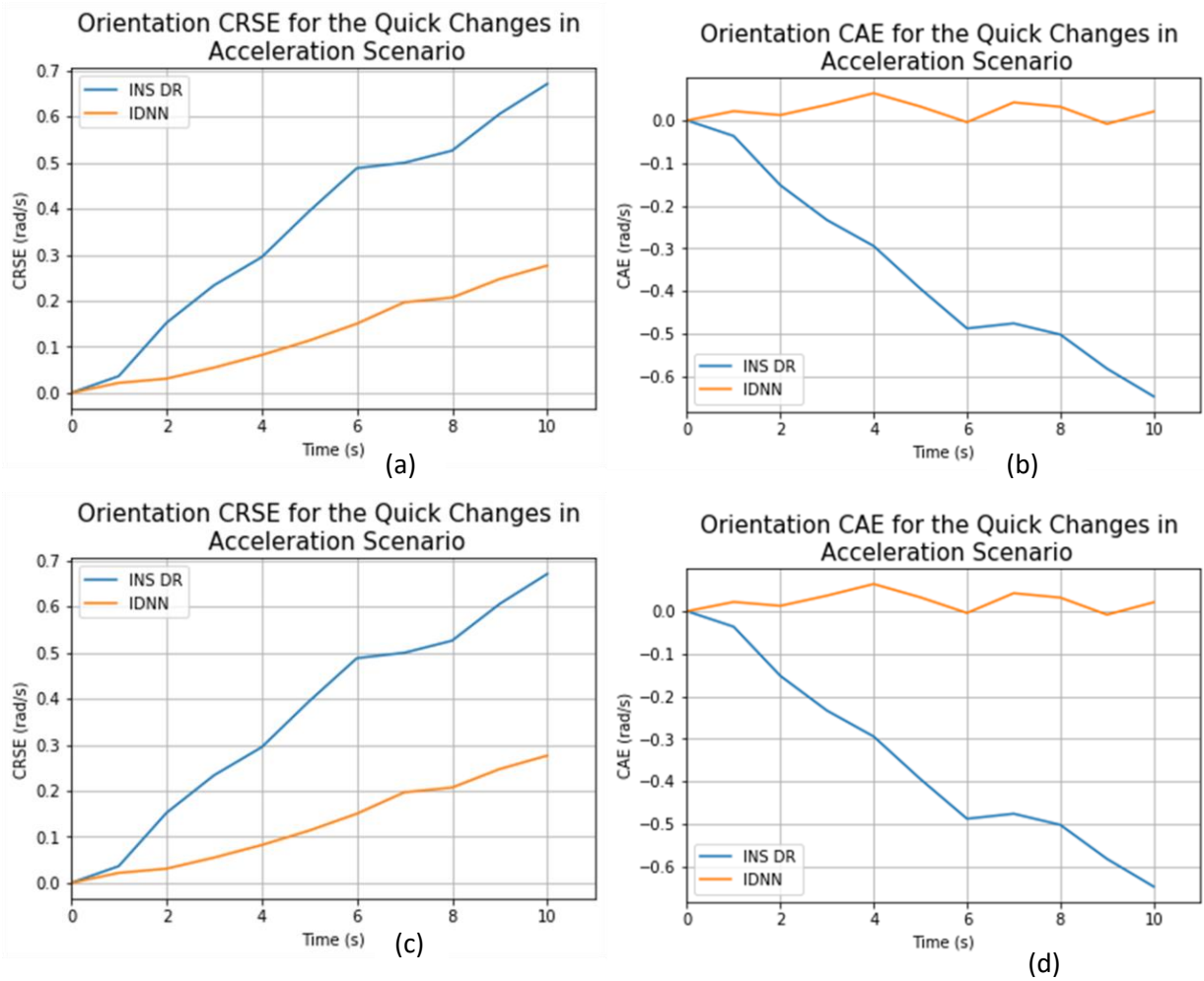

*Figure 7 showing the evolution of the estimation error over time in the quick changes in acceleration scenario based on the (a) displacement CRSE, (b) displacement CAE, (c) orientation rate CRSE and (d) orientation rate CAE*

## *4.4 Hard Brake Scenario*

The performance of the NN over the INS in the hard-brake scenario is evaluated over 14 test sequences averaging 188 m of travel with a 259 m maximum journey length. From Table 11, we observe that much to our expectations, the hard brake scenario proves to be more of a challenge for the accelerometer than the gyroscope as the INS struggles to accurately estimate the displacement and orientation rate of the vehicle within the simulated GPS outage period. As further emphasized on Figures 14-16, the NN significantly outperforms the INS DR across all performance metrics employed by a max and average CRSE of 15.80 m, 0.25 rad/s and 6.82 m, 0.09 rad/s for the NN compared to 133.12 m, 1.89 rad/s and 41.07 m and 0.37 rad/s respectively of the INS DR. The reliability of the NN in consistently correcting the INSs estimations to such accuracy is further established by its $\sigma$ value of 4.23 and 0.48. An example distribution of the accelerometers signal during this scenario is revealed on Figure 10.

*Table 11 showing the performance of the IDNN and INS DR on the hard brake scenario*

|  |  | IDNN (m) | INS DR (m) | IDNN (rad/s) | INS DR (rad/s) |
|---|---|---|---|---|---|
| **CRSE** | max | 15.80 | 133.12 | 0.25 | 1.89 |
|  | min | 1.15 | 5.74 | 0.03 | 0.06 |
|  | $\mu$ | 6.82 | 41.07 | 0.09 | 0.37 |
|  | $\sigma$ | 4.23 | 33.75 | 0.08 | 0.48 |
| **CAE** | max | 14.75 | 133.12 | 0.05 | 1.17 |
|  | min | 0.08 | 1.37 | 0.00 | 0.01 |
|  | $\mu$ | 0.44 | 26.50 | 0.02 | 0.21 |
|  | $\sigma$ | 3.89 | 34.70 | 0.01 | 0.33 |
| **AEPS (/s)** | max | 0.21 | 1.97 | 0.00 | 0.01 |
|  | min | 0.01 | 0.01 | 0.00 | 0.00 |
|  | $\mu$ | 0.06 | 0.47 | 0.00 | 0.00 |
|  | $\sigma$ | 0.06 | 0.43 | 0.00 | 0.00 |
| **Total Distance Covered** | max | 258.79 |  |  |  |
|  | min | 73.39 |  |  |  |
|  | $\mu$ | 188.48 |  |  |  |
| **Number of Sequences evaluated** |  | 14 |  |  |  |

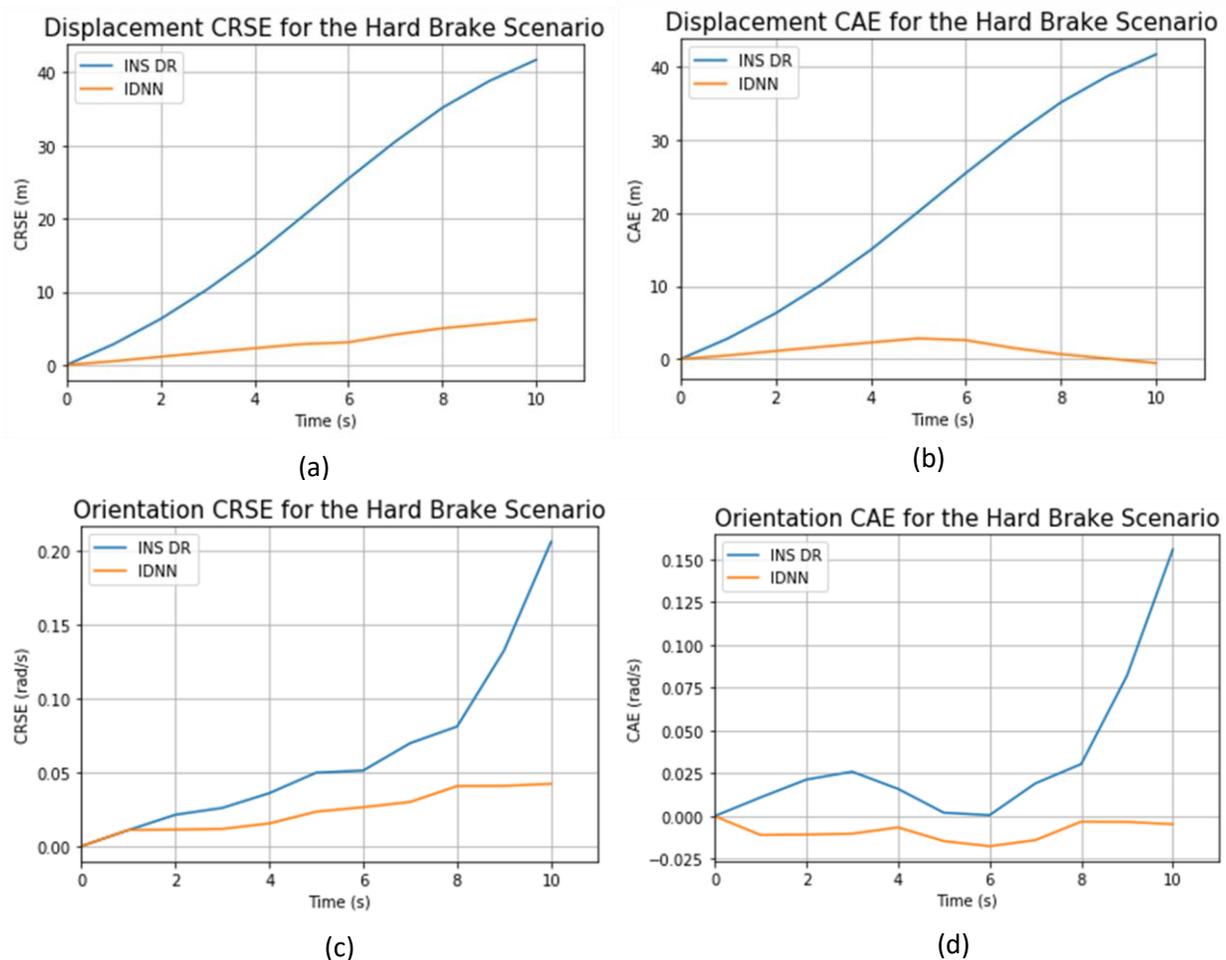

*Figure 8 showing the evolution of the estimation error over time in the hard brake scenario based on the (a) displacement CRSE, (b) displacement CAE, (c) orientation rate CRSE and (d) orientation rate CAE*

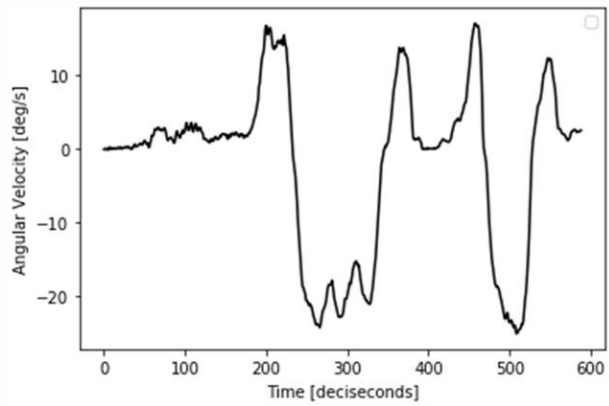 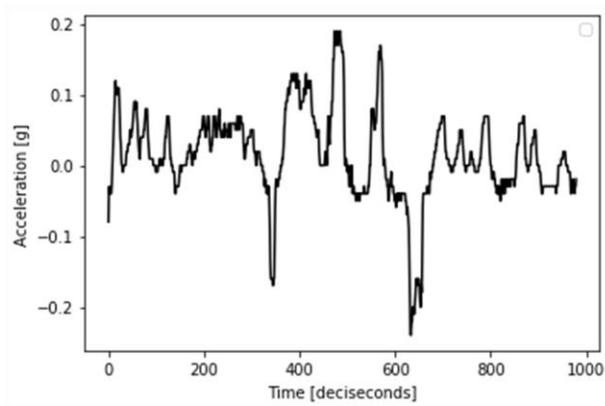

(a) (b)

*Figure 9 showing the variations in the (a) vehicle's angular velocity in the roundabout scenario and (b) vehicle's acceleration in the quick changes in acceleration scenario*

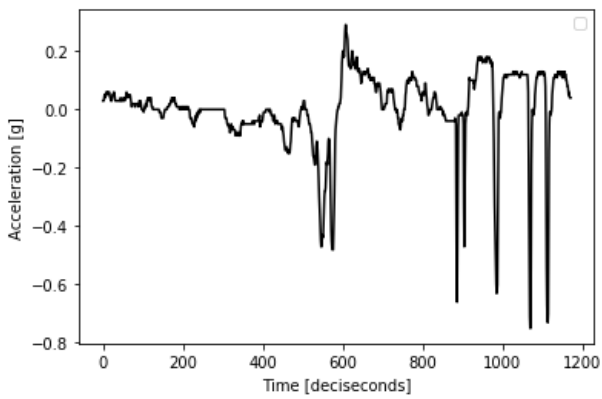 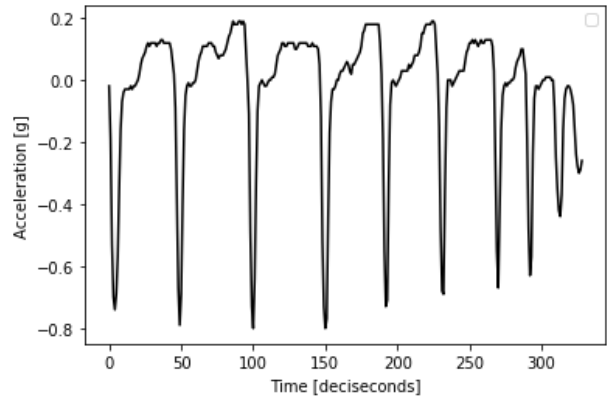

*Figure 10 showing the variations in the vehicle's acceleration in the hard brake scenario*

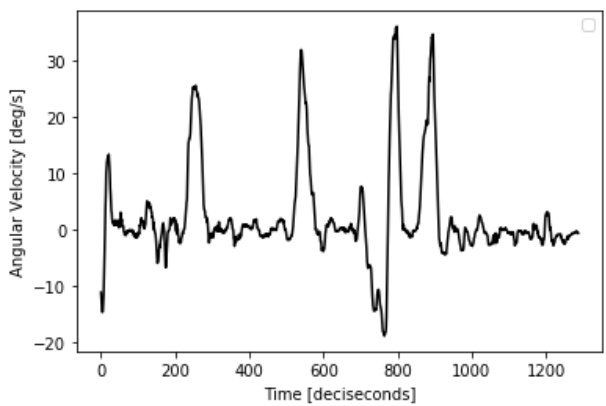 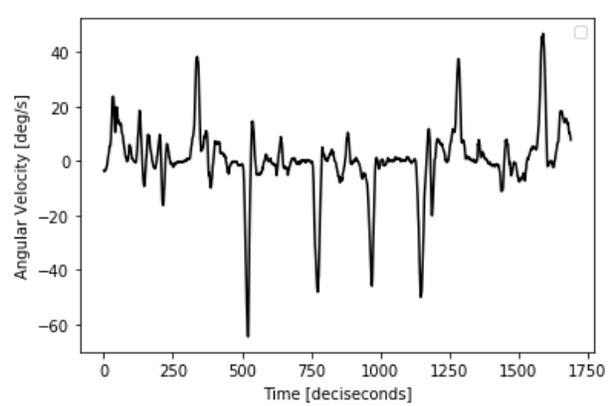

*Figure 11 showing the variations in the vehicle's angular velocity in the sharp cornering and successive left-right turns scenario.*

## 4.5 Sharp Cornering and Successive left-right turns Scenario

The sharp cornering and successive right-left turns scenario appears to be one of the most challenging for the INS on the CAE metric (see Figures 13-16). This scenario investigation involves an analysis on 40 test sequences over a maximum travel distance of approximately 110m. Reporting on the results presented on the Table 12 it can be observed that the INS has a maximum CRSE and CAE displacement of 92.06 m compared to 142.71 m and 8.49 m respectively of the NN. On the orientation rate the NN performs significantly better than the INS with a maximum CRSE and CAE of 0.41 rad/s and 0.13 rad/s against the INS's performance of 4.29 rad/s and 3.47 rad/s. These results further highlight the capability of the NN to significantly improve vehicular localisation during GPS outages with its reliability assured by its relatively low standard deviation. An example trajectory of the vehicle during the sharp cornering and successive left-right turn is shown on Figure 12. Figure 11 shows sample distributions of the gyroscopes measurement over time.

*Table 12 showing the performance of the IDNN and INS DR on the sharp cornering and successive left right-turns scenario*

|  |  | IDNN (m) | INS DR (m) | IDNN (rad/s) | INS DR (rad/s) |
|---|---|---|---|---|---|
| **CRSE** | max | 12.71 | 92.06 | 0.41 | 4.29 |
|  | min | 1.43 | 5.20 | 0.06 | 0.19 |
|  | $\mu$ | 6.77 | 39.35 | 0.19 | 1.99 |
|  | $\sigma$ | 2.83 | 26.91 | 0.09 | 1.42 |
| **CAE** | max | 8.49 | 92.06 | 0.13 | 3.47 |
|  | min | 0.10 | 2.02 | 0.00 | 0.01 |
|  | $\mu$ | 0.01 | 11.34 | -0.01 | -0.07 |
|  | $\sigma$ | 2.29 | 28.84 | 0.04 | 2.03 |
| **AEPS (/s)** | max | 0.25 | 0.94 | 0.02 | 0.16 |
|  | min | 0.00 | 0.00 | 0.00 | 0.00 |
|  | $\mu$ | 0.06 | 0.38 | 0.00 | 0.02 |
|  | $\sigma$ | 0.07 | 0.30 | 0.00 | 0.04 |
| **Total Distance Covered** | max | 109 |  |  |  |
|  | min | 21 |  |  |  |
|  | $\mu$ | 75 |  |  |  |
| **Number of Sequences evaluated** |  | 40 |  |  |  |

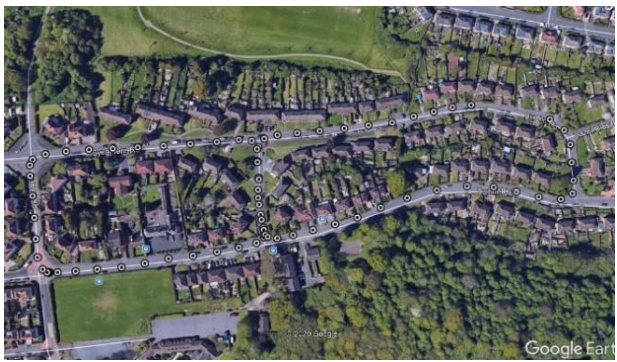
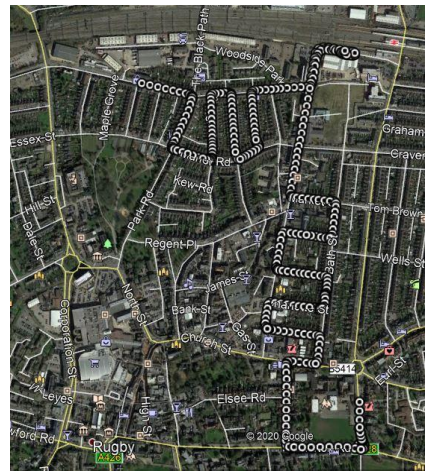

*Figure 12 Trajectory of V-Vw8 sharp cornering and successive left and right-turns data subset of the IO-VNBD*

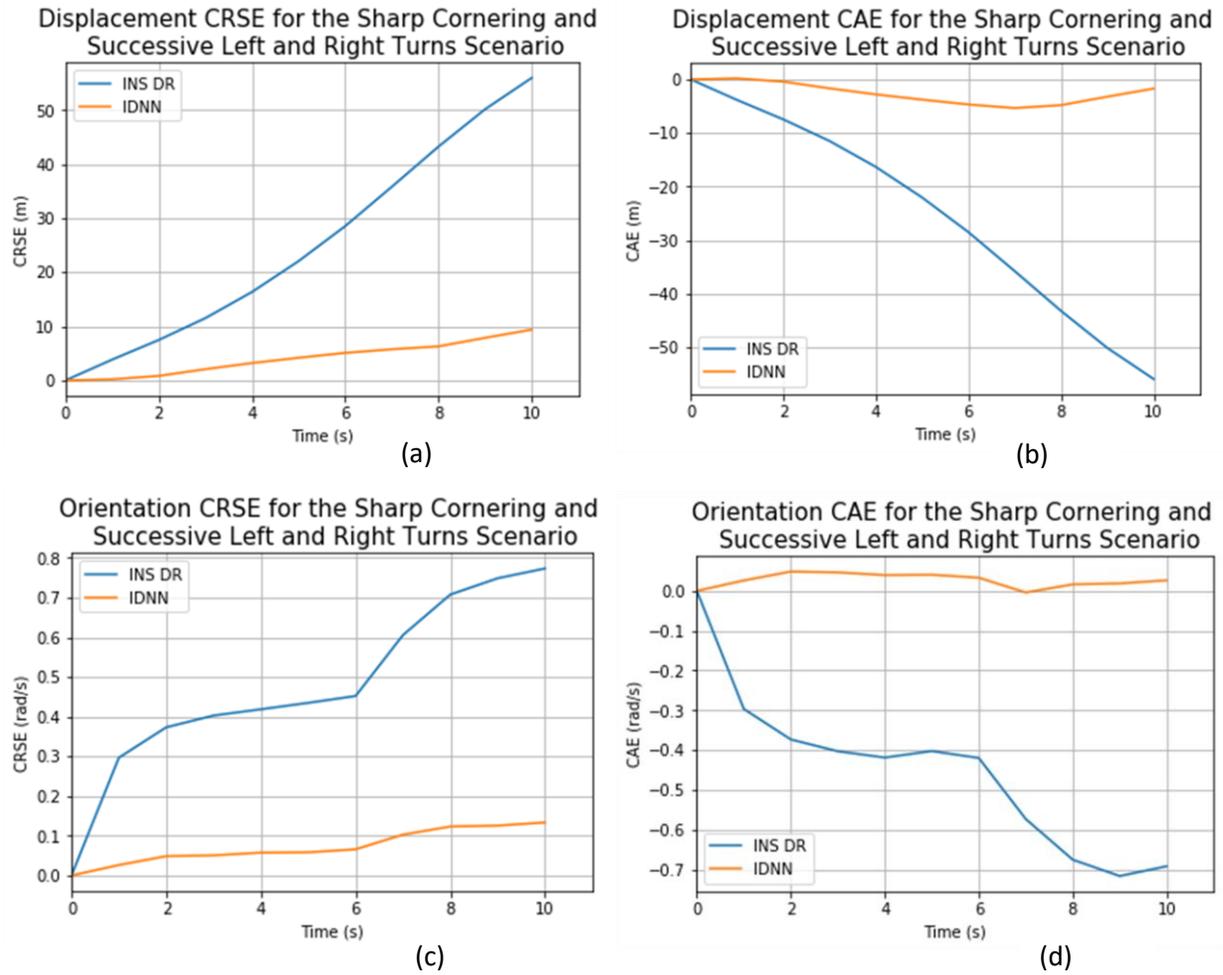

*Figure 13 showing the evolution of the estimation error over time in the hard brake scenario based on the (a) displacement CRSE, (b) displacement CAE, (c) orientation rate CRSE and (d) orientation rate CAE*

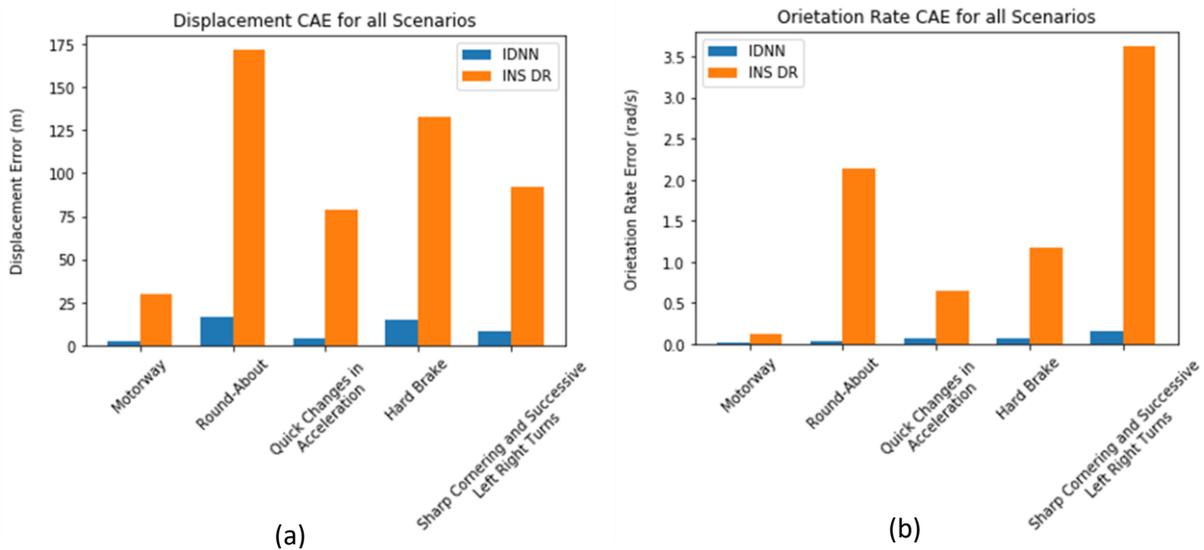

*Figure 14 showing the comparison of the CAE performance of the IDNN and INS DR across all investigated scenarios on the (a) displacement estimation and (b) orientation rate estimation*

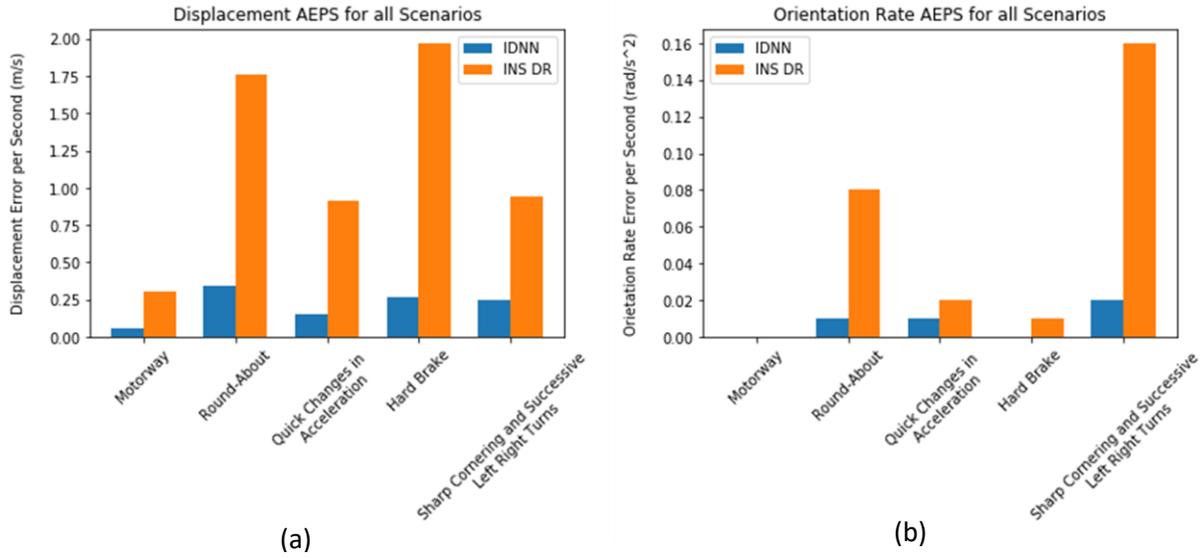

(a)                  (b)

*Figure 15 showing the comparison of the AEPS performance of the IDNN and INS DR across all investigated scenarios on the (a) displacement estimation and (b) orientation rate estimation*

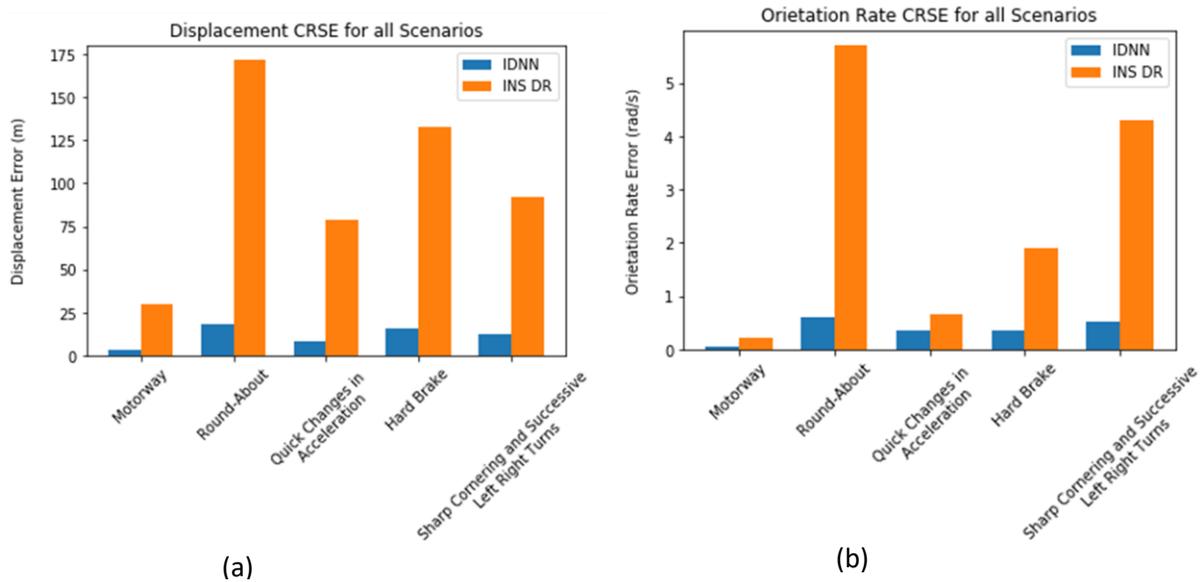

(a)                  (b)

*Figure 16 showing the comparison of the CRSE performance of the IDNN and INS DR across all investigated scenarios on the (a) displacement estimation and (b) orientation rate estimation*

## 5. *Conclusion*

We propose a Neural Network based approach inspired by the operation of the feedback control system to improve the localisation of autonomous vehicles and robots alike in challenging GPS deprived environments. The proposed approach is analytically compared to the INS specifically in scenarios characterised by hard braking, roundabouts, quick changes in vehicle acceleration, motorway, sharp cornering and successive left and right turns. By estimating the displacement and orientation rate of the vehicle within a GPS outage period, we show that the Neural Network based positioning approach outperforms the INS significantly in all investigated scenarios by providing up to 89.55% improvement on the displacement estimation and 93.35 % on the orientation rate estimation.

Nevertheless, we encounter the problem of model generalisation due to the varying characteristics of the sensor noise and bias in different journey domains as well as slight variations in the vehicular environment, trajectory and dynamics. These factors cause discrepancies between the training data and test data hindering better estimations. There is therefore the need to create a model capable of accounting for the variations in the sensors characteristics and environments towards the end purpose of robustly and accurately tracking the motion of the vehicle in various terrains. This will be the subject of our future research.